\documentclass[twocolumn]{el-author}

\newcommand{\ua}{\uparrow}
\newcommand{\nc}{\newcommand}
\nc{\da}{\downarrow} \nc{\hc}{\hat{c}} \nc{\hS}{\hat{S}}
\nc{\bra}{\langle} \nc{\ket}{\rangle} \nc{\eq}{equation (\ref}
\nc{\h}{\hat} \nc{\hT}{\h{T}}\nc{\be}{\begin{eqnarray}}
\nc{\ee}{\end{eqnarray}}\nc{\rd}{\textrm{d}}\nc{\e}{eqnarray}\nc{\hR}{\hat{R}}\nc{\Tr}{\mathrm{Tr}}
\nc{\tS}{\tilde{S}}\nc{\tr}{\mathrm{tr}}\nc{\8}{\infty}\nc{\lgs}{\bra\ua,\phi|}\nc{\rgs}{|\ua,\phi\ket}
\nc{\hU}{\hat{U}}\nc{\lfs}{\bra\phi|}\nc{\rfs}{|\phi\ket}\nc{\hZ}{\hat{Z}}\nc{\hd}{\hat{d}}\nc{\mD}{\mathcal{D}}
\nc{\bd}{\bar{d}}\nc{\bc}{\bar{c}}\nc{\mc}{\mathcal}\nc{\ea}{eqnarray}\nc{\mG}{\mathcal{G}}\nc{\bce}{\begin{center}}
\nc{\ece}{\end{center}}
\date{xxx 2014}

\usepackage[width=18.46cm,height=10in]{geometry}
\usepackage[noadjust]{cite}
\usepackage{srcltx}
\usepackage{psfrag}
\usepackage{epsfig}
\usepackage{amsmath}

\begin{document}

\title{The 2D Linearly Polarized Near-Field Focusing Based on Angularly Discretized Slot Arrays}

\author{Menglin Chen, Shulabh Gupta, Zilong Ma and Lijun Jiang}

\abstract{A 2-D near-field focusing design is proposed based on the circular slot array waveguide structures, synthesized using the array-factor theory, and demonstrated by full-wave simulations. The principle of beam-focusing is extended to the 2-D angularly discretized configuration using regular center-fed linear slots arranged in a circular pattern. By the mirror image arrangement of the slots, a linearly polarized focus in the near-field of the antenna, with negligible cross-polarization is achieved. Its beam-focusing properties are discussed in details and demonstrated by simulations.}

\maketitle

\section{Introduction}

Recently, there has been an increased interest in the near-field focusing at microwave and millimetre wave frequencies where the radiated power is focused at a desired location close to the antenna. They have several applications in the field of RFID, imaging and various sensing systems\cite{app_microstrip_array,app_rfid}. Such a focusing system typically requires dispersion engineered structures in order to control the phase of various radiating components. Various works have been reported in literature, where the beam focusing is achieved using printed antennas, such as patch arrays \cite{app_microstrip_array,app_rfid}, and leaky wave antennas \cite{Caloz1, slot_lwl, Part1, Width_tapered_lwl}. Another attractive implementation of such focusing structure is the slot array waveguide. Compared to conventional patch array focusing structure which requires a complicated corporate feeding networks and the printed leaky-wave antennas, slot waveguides can be conveniently fed by coaxial type excitations, and are particularly suitable for high frequency applications due to low dielectric and conductor losses \cite{Ando1, Ando2}. In particular, a beam-focusing antenna with circular polarization has been shown for a 2-D slot aperture antenna in \cite{Part1, Part2}, and one with the linear polarization in \cite{conf_lwl}.

In this work, we propose a \emph{2-D near-field focusing based on the slot array configuration} to achieve a \emph{linearly polarized} focus in the near-field of the radiating aperture with the following two features: 1) A \emph{pure linearly polarized focal beam, with negligible cross-polarization}, using the mirror image arrangement of the slots, 2) \emph{Co-linear polarization}, parallel to the 2D slot array structure.

\section{The Lensing Formulation}

\begin{figure}[htbp]
\begin{center}
\psfrag{a}[c][c][0.8]{$z_0$}
\psfrag{b}[c][c][0.8]{$-\ell_0$}
\psfrag{c}[c][c][0.8]{$+\ell_0$}
\psfrag{d}[c][c][0.8]{$\ell_{m+1}$}
\psfrag{e}[c][c][0.8]{$\ell_m$}
\psfrag{y}[c][c][0.8]{$y$}
\psfrag{z}[c][c][0.8]{$z$}
\includegraphics[width=\columnwidth]{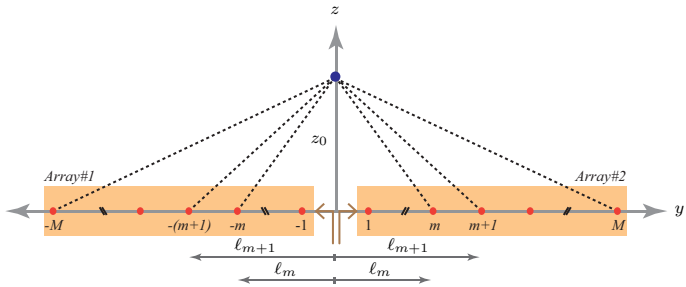}
\caption{An array of centre-fed non-uniformly spaced ideal radiators, to focus a beam at a distance $z_0$.}\label{Fig:IdealModel}
\end{center}
\end{figure}

Consider an array of non-uniformly spaced ideal point radiators symmetrically located around the $z-$axis, at $y = \pm\ell_m$ where $m\in[1, M]$, as shown in Fig.~\ref{Fig:IdealModel}. The objective is to find exact locations of these point radiators so that the individual radiation contributions from the radiators converge to a single point at $z = z_0$, resulting in a beam-focusing effect \cite{Caloz2}. 

\vspace{-0.5cm}

Let's assume that point radiators are uniformly excited at $f_0$, and the feed location is the phase reference. In order to achieve beam focus, all waves emanating from point radiators must add in phase at $z=z_0$. Mathematically, $\phi_{m+1} - \phi_m = 2n\pi$ $\forall m$, where $\phi_m$ is the phase at the desired focal point generated by the $m^\text{th}$ radiator and $n$ is any integer. $\phi_m$  consists of two parts: $\phi^\text{(1)}_m$, the phase difference between the excitation point and radiator; and $\phi^\text{(2)}_m$, the phase delay between the radiator and the desired focal point. The first contribution can be conveniently written in terms of the guided-wavelength $\lambda_g$, to model a general configuration of the series fed array. Specifically, for the $m^\text{th}$ radiator, $\phi^\text{(1)}_m = 2\pi\ell_m/\lambda_g$, assuming the slot represents a small perturbation in the waveguide. The second part $\phi^\text{(2)}_m$ can be written as: $\phi^\text{(2)}_m = 2\pi d_m/\lambda_0$, where $d_m$ is the distance between the $m^\text{th}$ radiator and the focal point, and $\lambda_0$ is the free-space wavelength at $f_0$. The total phase shift $\phi_m = (2\pi/\lambda_g)\ell_m + (2\pi/\lambda)\sqrt{z_0^2 + \ell_m^2}$. Finally, using the above equation for the $m^\text{th}$ and the $(m+1)^\text{th}$ radiators in the phase matching condition, a quadratic equation in $\ell_{m+1}$ is obtained as $ \left(1/\lambda_g^2 - 1/\lambda^2\right)\ell_{m+1}^2 - (2R_m/\lambda_g)\ell_{m+1} +\left(R_m^2-z_0^2/\lambda^2 \right) = 0$, where $R_m = 1 + \ell_m/\lambda_g + (1/\lambda)\sqrt{z_0^2 + \ell_m^2}$. Here, to minimize the total length of the array and minimize grating lobes, $n$ is chosen to be one. Finally, using the position of the first radiator, those for remaining radiators $\ell_{m+1}$ can be solved iteratively.


\begin{figure}[htb]
\begin{center}
\psfrag{a}[l][c][0.5]{$s_0$}
\psfrag{b}[l][c][0.5]{$w_0$}
\psfrag{d}[c][c][0.5]{$l_0$}
\psfrag{c}[c][c][0.8]{$\ell_1$}
\psfrag{f}[c][c][0.8]{$l_s$}
\psfrag{e}[c][c][0.8]{$w$}
\psfrag{h}[c][c][0.8]{$\Delta\theta = 36^\circ$}
\psfrag{g}[c][c][0.8]{$l_s$}
\psfrag{i}[c][c][0.8]{T-junction}
\psfrag{j}[c][c][0.7]{\shortstack{$1:N$ power\\ divider}}
\psfrag{k}[c][c][0.8]{$r$}
\psfrag{l}[c][c][0.8]{0}
\psfrag{m}[c][c][0.8]{$w$}
\psfrag{n}[c][c][0.8]{short}
\psfrag{x}[c][c][0.8]{$x$}
\psfrag{y}[c][c][0.8]{$y$}
\psfrag{z}[c][c][0.8]{$z$}
\psfrag{w}[r][c][0.8]{infinite PEC}
\psfrag{v}[c][c][0.8]{$\ell_t$}
\includegraphics[width=0.8\columnwidth]{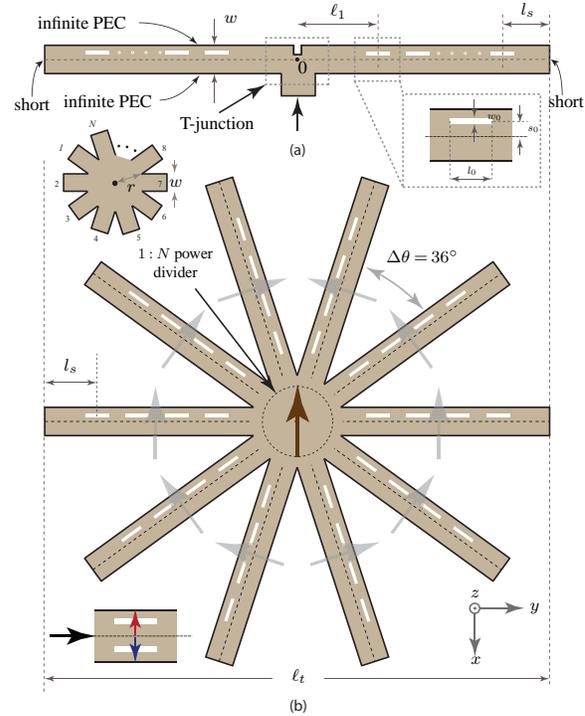}
\caption{Illustration of the centre-fed slot array configuration. a) Linear slot array. b) 2-D slot array. The parameters are:  $f_0=10$~GHz, $w = 12.2$~mm, $\lambda_g = 24.6$~mm, $M = 10$, $z_0 = 10$~cm, $l_1 = 25$~mm, $s_0 = 0.425$~mm, $w_0 = 0.4$~mm, $l_0 = 11$~mm, $l_s = 3/4 \lambda_g$, and total length of the structure $\ell_t = 36.84$~cm. Substrate Taconic RF-30 substrate, thickness $h=2.29$~mm, $\varepsilon_r=3.0$.} \label{Fig:slot_structure}
\end{center}
\end{figure}

\vspace{-1cm}

\section{1-D Near-Field Focusing using Centre-fed Linear Slot Array}
The ideal radiator array can be conveniently implemented by using the \emph{slot array waveguide}. The layout is shown in Fig.~\ref{Fig:slot_structure}(a), where the two halves of the structure are fed using a waveguide T-junction \cite{T-junction}. The dimensions of the waveguide are chosen to ensure that the design frequency $f_0$ is larger than its cut-off frequency. To operate the waveguide in the resonant mode, both ends of the waveguide are terminated with a shorting wall. Furthermore, infinite perfect electric conductors (PEC) are assumed at $x = \pm w/2$. 

\begin{figure}[htbp]
\begin{center}
\psfrag{a}[c][c][0.8]{$y$~(mm)}
\psfrag{b}[c][c][0.8]{$z$~mm}
\psfrag{c}[c][c][0.7]{power density in log scale~(W/m$^2$)}
\psfrag{d}[l][c][0.7]{HFSS}
\psfrag{e}[l][c][0.7]{Analytical}
\psfrag{f}[l][c][0.7]{uniform}
\psfrag{g}[l][c][0.7]{lens}
\psfrag{h}[c][c][0.7]{$\Delta z = 60$~mm}
\psfrag{x}[l][c][0.7]{(a)}
\psfrag{y}[l][c][0.7]{(b)}
\includegraphics[width=\columnwidth]{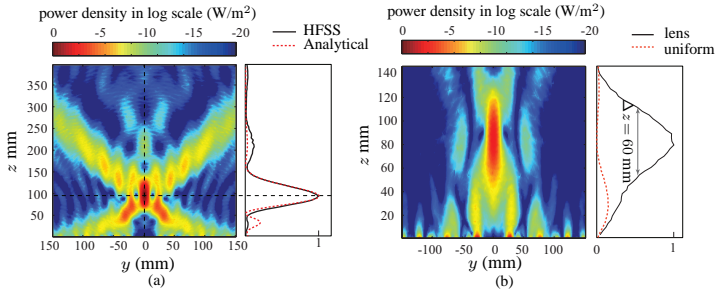}
\caption{HFSS simulated power density distribution in the $y-z$ planes of a) a 1-D linear slot array of Fig.~\ref{Fig:slot_structure}(a), and b) a 2-D slot array of  Fig.~\ref{Fig:slot_structure}(b) with the mirror image arrangements.} 
\label{Fig:Linear_Fields}
\end{center}
\end{figure}

Fig.~\ref{Fig:Linear_Fields}(a) shows the typical full-wave simulated power density distribution above the slot array, for the case of $z_0=10$~cm, to illustrate the beam focussing. The physical parameters of the waveguide and the $\lambda_g$-spaced radiating slots are given in Fig.~\ref{Fig:slot_structure}, with the locations of the slots calculated using the array factor approach explained above. As desired, the power density is maximum at the desired focal point. Furthermore, the power density profile along the $z-$axis show an excellent agreement with that of the ideal radiators. It is noted that in this example, the beam is focused in the reactive near-field region of the antenna structure, i.e. $z_0 < 0.62\sqrt{D^3/\lambda}$ \cite{reactive_region}, where $D$ is the size of the radiating aperture.

\section{2-D Near-Field Focusing using Center-fed 2-D Slot Array}

The linear slot array can only focus a beam in the $y-z$ plane only. In order to focus a beam into a 3-D region of the free space above the structure, a 2-D radiating aperture is required. Based on the linear structure in Fig.~\ref{Fig:slot_structure}(a), a 2-D structure can be formed as shown in Fig.~\ref{Fig:slot_structure}(b), where the linear slot array is rotated around its centre to form a star-shape circular configuration. Such a structure can be fed by a vertical monopole at the centre using a $1:N$ waveguide power divider \cite{power_divider}.

The linear slot array is linearly polarized along $x-$axis, i.e. $E_x$ is dominant. To achieve the same polarization in the 2-D structure, the left and right-half of the 2D slot array should be the mirror image of each other to cancel the horizontal $E_y$ polarization. This can be achieved by placing the longitudinal slots in the appropriate halves of the individual waveguide arms, as shown in Fig.~\ref{Fig:slot_structure}(b), which results in an overall $x-$directed polarization, due to the mirror image design.

\begin{figure}[htbp]
\begin{center}
\psfrag{x}[c][c][0.8]{$x$~(mm)}
\psfrag{y}[c][c][0.8]{$y$~(mm)}
\psfrag{e}[l][c][0.7]{$\Delta y = 17$~mm}
\psfrag{f}[c][c][0.7]{power density in log scale~(W/m$^2$)}
\psfrag{b}[c][c][0.8]{$E$ field (V/m)}
\psfrag{c}[l][c][0.8]{$E_x$}
\psfrag{d}[l][c][0.8]{$E_y$}
\psfrag{h}[l][c][0.7]{uniform}
\psfrag{g}[l][c][0.7]{lens}
\includegraphics[width=\columnwidth]{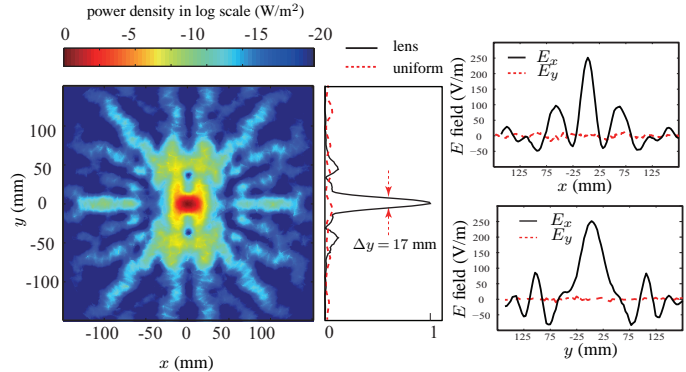}
\caption{Power density distribution in the focal plane ($z=10$~cm) of the 2D slot array of Fig.~\ref{Fig:slot_structure}(b), along with its two orthogonally polarized $E_x$ and $E_y$ components along $x-$ and $y-$axes.} 
\label{Fig:FieldPol2D}
\end{center}
\end{figure}

Fig.~\ref{Fig:Linear_Fields}(b) shows the full-wave simulated power density distribution for the 2-D focusing case, demonstrating a strong beam focus at $z=10$~cm, in one of the longitudinal planes. Fig.~\ref{Fig:FieldPol2D} shows the power density distribution along the transverse focal plane, parallel to the structure. A spot size with a 3-dB beamwidth of $0.8\lambda_0$ and $0.56\lambda_0$ along $x-$ and $y-$directions, respectively is achieved in this design. Fig.~\ref{Fig:FieldPol2D} also shows the two orthogonal polarizations $E_x$ and $E_y$ in the focal plane along $x-$ and $y-$direction. A \emph{dominant $E_x$ polarization} is observed along both axes, thereby validating the usage of yhr mirror image design to obtain a linearly polarized beam-focusing.

\begin{figure}[htbp]
\begin{center}
\psfrag{a}[c][c][0.8]{vertical distance $z$~(mm)}
\psfrag{b}[c][c][0.8]{power density~(W/m$^2$)}
\psfrag{c}[l][c][0.8]{10 arms}
\psfrag{d}[l][c][0.8]{14 arms}
\psfrag{e}[l][c][0.8]{(a)}
\psfrag{f}[l][c][0.8]{(b)}
\includegraphics[width=0.7\columnwidth]{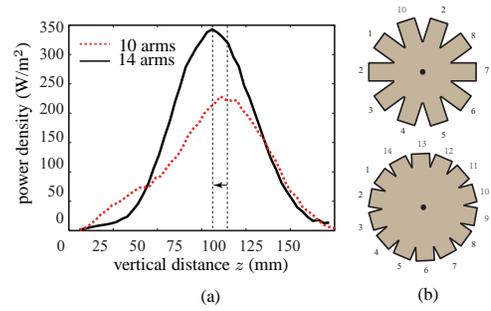}
\caption{Power density comparison by increasing the no. of branches. a) power density along $z$ axis. b) illustration of the corresponding power dividers.} 
\label{Fig:Parametric}
\end{center}
\end{figure}

The proposed 2-D slot array employs a power divider to distribute the input power into $N$ discrete channels. Ideally the entire 2-D aperture should radiate and contribute to the focusing. However, the power division scheme, may be seen as angularly discretizing the physical aperture of the slot array. To study the effect of this discretization, two 2-D slot arrays were simulated for the case of $1:10$ and $1:14$ power dividers. Fig.~\ref{Fig:Parametric} shows the field power density variation perpendicular to the 2-D array along $z$.  As expected, larger number of wave guiding channels results in an increased power density at the focal point by over 50\%. Moreover, the focal length converges towards the design value of $z_0=10$~cm, as the number of waveguide branches is increased, which is the result of better aperture sampling.

\section{Conclusions}
A 2-D linearly polarized near-field focusing is proposed based on an angularly discretized slot array waveguide and demonstrated. The lensing formulation employing the array synthesis process has been applied to compute locations of discrete radiating elements to achieve a focus at a given distance from the the 1-D antenna aperture in its near field. The method is then applied to a 2-D configuration in the angularly discretized slot array. Exploiting the mirror image configuration of the 2-D array, a linearly polarized beam focused in the near field of the antenna aperture with negligible cross-polarization has been achieved.

\vspace{-0.75cm}
\vskip3pt
\ack{This work was supported in part by HK ITP/026/11LP, HK GRF 711511, HK GRF 713011, HK GRF 712612, and  NSFC 61271158.}

\vskip5pt

\noindent Menglin Chen, Shulabh Gupta, Zilong Ma and Lijun Jiang (\textit{The University of Hong Kong, HK}), E-mail: menglin@connect.hku.hk
\vskip3pt


\bibliographystyle{IEEEtran}
\bibliography{References_ME}

\begin{thebibliography}{10}
\providecommand{\url}[1]{#1}
\csname url@samestyle\endcsname
\providecommand{\newblock}{\relax}
\providecommand{\bibinfo}[2]{#2}
\providecommand{\BIBentrySTDinterwordspacing}{\spaceskip=0pt\relax}
\providecommand{\BIBentryALTinterwordstretchfactor}{4}
\providecommand{\BIBentryALTinterwordspacing}{\spaceskip=\fontdimen2\font plus
\BIBentryALTinterwordstretchfactor\fontdimen3\font minus
  \fontdimen4\font\relax}
\providecommand{\BIBforeignlanguage}[2]{{%
\expandafter\ifx\csname l@#1\endcsname\relax
\typeout{** WARNING: IEEEtran.bst: No hyphenation pattern has been}%
\typeout{** loaded for the language `#1'. Using the pattern for}%
\typeout{** the default language instead.}%
\else
\language=\csname l@#1\endcsname
\fi
#2}}
\providecommand{\BIBdecl}{\relax}
\BIBdecl

\bibitem{app_microstrip_array}
M.~Bogosanovic and A.~G. Williamson, ``Microstrip antenna array with a beam
  focused in the near-field zone for application in noncontact microwave
  industrial inspection,'' \emph{IEEE Trans. Instrumentation and Measurement},
  vol.~56, no.~6, pp. 2186--2195, Dec. 2007.

\bibitem{app_rfid}
A.~Buffi, A.~A. Serra, P.~Nepa, H.~T. Chou, and G.~Manara, ``A focused planar
  microstrip array for 2.4 ghz rfid readers,'' \emph{IEEE Trans. Antennas
  Propagat.}, vol.~58, no.~5, pp. 1536--1543, May 2010.

\bibitem{Caloz1}
C.~Caloz and T.~Itoh, ``Electromagnetic metamaterials, transmission line theory
  and microwave applications,'' \emph{Wiley-IEEE Press}, 2005.

\bibitem{slot_lwl}
A.~J. Martinez-Ros, F.~Quesada-Pereira, A.~Alvarez-Melcon, G.~Goussetis, A.~R.
  Weily, and Y.~J. Guo, ``Frequency steerable two dimensional focusing using
  rectilinear leaky-wave lenses,'' \emph{IEEE Trans. Antennas Propagat.},
  vol.~59, no.~2, pp. 407--415, Feb. 2011.

\bibitem{Part1}
J.~L. Gomez-Tornero, D.~Blanco, E.~Rajo-Iglesias, and N.~Llombart,
  ``Holographic surface leaky-wave lenses with circulary-polarized focused
  near-fields -- part 1: Concept, design and analysis theory,'' \emph{IEEE
  Trans. Antennas Propagat.}, vol.~61, no.~7, pp. 3475--3485, Jul. 2013.

\bibitem{Width_tapered_lwl}
A.~J. Martinez-Ros, J.~L. Gomez-Tornero, F.~J. Clemente-Fernandez, and
  J.~Monzo-Cabrera, ``Microwave near-field focusing properties of width-tapered
  microstrip leaky-wave antenna,'' \emph{IEEE Trans. Antennas Propagat.},
  vol.~61, no.~6, pp. 2981--2990, Jun. 2013.

\bibitem{Ando1}
J.~H. Lee, J.~Hirokawa, and M.~Ando, ``A center-feed waveguide transverse slot
  linear array using a transverse-slot feed for blocking reduction,''
  \emph{IEICE Trans. on Communications}, vol. E94-B, no.~1, pp. 326--329, Jan.
  2011.

\bibitem{Ando2}
M.~Takahashi, J.~Takada, M.~Ando, and N.~Goto, ``A slot design for uniform
  aperture field distribution in single-layered radial line slot antennas,''
  \emph{IEEE Trans. Antennas Propag.}, vol.~39, no.~7, pp. 954--959, July.
  1991.

\bibitem{Part2}
J.~L. Gomez-Tornero, D.~Blanco, E.~Rajo-Iglesias, and N.~Llombart,
  ``Holographic surface leaky-wave lenses with circulary-polarized focused
  near-fields -- part 2: Experiments and description of frequency steering of
  focal length,'' \emph{IEEE Trans. Antennas Propagat.}, vol.~61, no.~7, pp.
  3486--3494, Jul. 2013.

\bibitem{conf_lwl}
J.~L. Gomez-Tornero, A.~J. Martinez-Ros, N.~Llombart, D.~Blanco, and
  E.~Rajo-Iglesias, ``Near-field focusing with holographic two-dimensional
  tapered leaky-wave slot antennas,'' \emph{Antenna and Propagation Conference
  (EuCAP), 2012 6th European}, pp. 234--238, Mar. 2012.

\bibitem{Caloz2}
I.~H. Lin, C.~Caloz, and T.~Itoh, ``Near-field focusing by a nonuniform
  leaky-wave interface,'' \emph{Microw. Opt. Technolgoy Lett.}, vol.~44, no.~5,
  pp. 416--418, Mar. 2005.

\bibitem{T-junction}
C.~G. Montgomery, R.~H. Dicke, and E.~M. Purcell, ``Principles of microwave
  circuits,'' \emph{New York: McGraw-Hill Book Company INC}, 1948.

\bibitem{reactive_region}
C.~A. Balanis, ``Antenna theory: Analysis and design,'' \emph{New York: Wiley},
  2005.

\bibitem{power_divider}
M.~E. Bialkowski and V.~P. Waris, ``Analysis of an n-way radial cavity divider
  with a coaxial central port and waveguide output ports,'' \emph{IEEE Trans.
  Microw. Theory Tech.}, vol.~44, no.~11, pp. 2010--2016, Nov. 1996.

\end{thebibliography}

\end{document}